# Dimensional effects in photoelectron spectra of Ag deposits on GaAs(110) surfaces

V. Gasparian*, E. Louis and J.A. Vergés+

*Departamento de Física Aplicada, Universidad de Alicante,
Apartado 99, E-03080 Alicante, Spain.*

It is shown that the peak structure observed in angle-resolved photoelectron spectra of metallic deposits can only be unambiguously associated to single electronic states if the deposit has a two dimensional character (finite along one spatial direction). In one and zero dimensions the density of states shows peaks related to bunches of single electron states (the finer structure associated to the latter may not always be experimentally resolved). The characteristics of the peak structure strongly depend on the band dispersion in the energy region where they appear. Results for the density of states and photoemission yield for Ag crystallites on GaAs(110) are presented and compared with experimental photoelectron spectra.

The increasing availability of synchrotron sources is allowing the accumulation of experimental information on quantum size effects in metallic deposits on either metal or semiconductor surfaces[1-3]. This is for instance the case of silver deposits on Cu or GaAs substrates. Ag overlayers grow epitaxially on a Cu(111) surface. Instead, Ag deposition on the GaAs(110) surface produce disordered films which, after annealing, give rise to Ag crystallites with a triangular cross section[4]. In both cases, Angularly Resolved Photoemission Electron Spectroscopy (ARPES) show well defined peaks whose separation seems to be independent of energy, and decreases with the amount of deposit.

In this work we show that these experimental data reveal important dimensional effects which have not been discussed previously. Although a full understanding of the features of the photoelectron spectra would require a complete analysis of the specific electronic structure of a given deposit, some general trends can be derived and those are the object of the present work. We shall start by discussing the features of the free electron density of states in finite systems in two, one and zero dimensions. Then we shall present numerical results for the density of states and the photoemission yield calculated by means of a model suited to the case of silver deposits on the GaAs (110) surface.

The density of states for free electrons in a $L \times M \times N$ cluster is given by (atomic units will be used, $\hbar = 2m = e = 1$),

$$\rho(E) = \sum_{l,m,n} \delta \left[ E - (\frac{\pi l}{L})^2 - (\frac{\pi m}{M})^2 - (\frac{\pi n}{N})^2 \right] \quad (1)$$

In the limit of an infinite solid the sums can be replaced by integrals giving the standard densities of states in one, two and three dimensions which show no oscillations[5]. Size effects can only be taken into account if these sums are not replaced by integrals. In two dimensions (a system finite in the $l$ direction) $m$ and $n$ are good quantum numbers and the oscillations in the density of states are due to single electronic states. This interpretation was offered in[1] to understand the origin of the quantum-well states observed in angle-resolved photoelectron spectra of Ag films epitaxially grown on Au(111). This is not the case, however, in one and zero dimensions. In one dimension (a cluster finite in the $l$ and $m$ directions), the sums can be carried out in the limit of large energies by using Poisson formula[6], leading to,

$$\rho_{1D}(E_n) \approx \frac{1}{2\pi} \left[ 1 + J_0(2L\sqrt{E_n}) + J_0(2M\sqrt{E_n}) + 2J_0(L\sqrt{E_n(L^2+M^2)}) \right] \quad (2)$$

where $E_n = E - (\pi n/N)^2$ and $J_0$ is the Bessel function of the first kind and zero order. A constant density of states is recovered for an infinite solid. The accuracy of this expression is illustrated in Fig. 1, which shows the results obtained from Eq. (2) and those resulting from a numerical evaluation of Eq. (1) for the case of a 20 × 20 cluster. It is noted that the peak structures are almost identical. However, whereas Eq. (2) gives an amplitude which decreases with energy, in the numerical results the amplitude increases. The reason is that, aiming to simulate the almost constant broadening in the experimental spectra, the numerical calculations were carried out by adding a constant imaginary part to the energy, whereas the quadratic dispersion relation would have required an imaginary part dependent on the energy.

The oscillations of Fig. 1 *are not related to single electron states*, but they are rather associated to several states each. It should be noted, however, that, depending upon experimental resolution, temperature and deposit dimensions, the finer structure associated to single electron states may also be observed. The separation between these oscillations (peaks) decreases with the cluster size as $1/L^2$. On the other hand, peak separation, $E_i - E_{i-1}$, is not constant but rather parabollic with increasing i. This is a consequence of the quadratic dispersion relation characteristic of free electrons, and should therefore strongly depend on the band dispersion in the energy region where the peaks lie (see below). It can be checked that if one assumes a linear dispersion relation (as can be found in some energy regions of actual band structures of many metals[1]) the density of states in a $L \times L$ cluster shows peaks separated by a constant energy, and whose weight increases linearly with energy (in an infinite system such dispersion relation would give a density of states proportional to E).

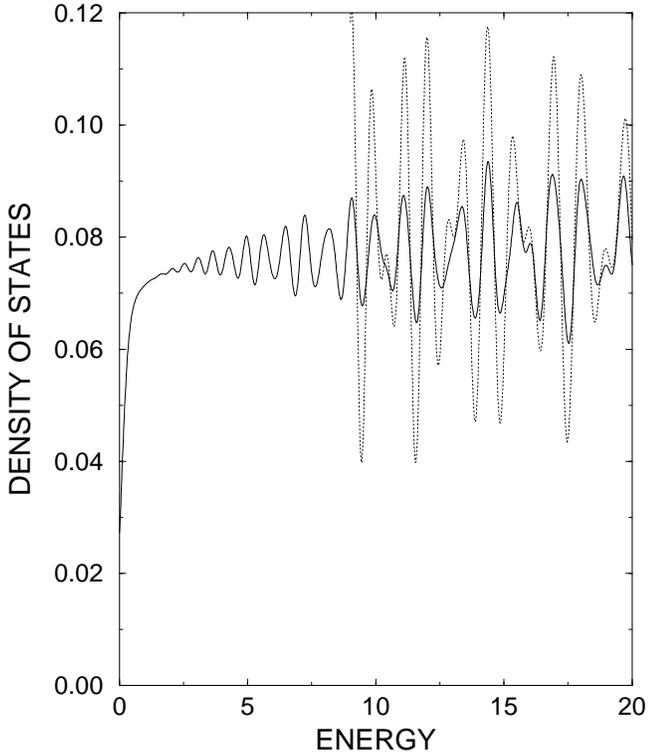

FIG. 1. Numerical (continuous line) and approximated (Eq. (2)) results for the total density of electron states in a $20 \times 20$ cluster calculated within the free electron approximation. Results obtained with the approximated expression are plotted only for large values of the arguments of the Bessel function in Eq. (2). Both the density of states and the energy are given in atomic units ($\hbar = 2m = e = 1$).

As outlined above, the Ag crystallites formed on GaAs(110) have not a square cross section but they rather have a triangular shape. It can be easily checked that the density of states in a triangle shows oscillations similar to those found in a square. To illustrate this point we discuss a simple case. The wave function within a triangular zone obtained by halving a square with infinite barriers at the boundaries is given by, $\Psi(\mathbf{r}) = C\left[\sin(k_x x)\sin(k_y y) - \sin(k_x y)\sin(k_y x)\right]$, C is a normalization constant. Note that the wave function within a square with infinite barriers at the boundaries can be expressed by either of the two terms in $\Psi$. The dispersion relation is that of Eq. (1) with the restriction that $l \neq m$. The resulting density of states is essentially that of Fig. 1.

These oscillations survive in the case of clusters in three dimensions. For instance, the density of states for a $L \times L \times N$ cluster can be approximated by means of tecniques similar to those used in 1D, giving,

$$\rho_{0D}(E) \approx \frac{1}{4\pi^2}\left(\sqrt{E} + \frac{\sin(2N\sqrt{E})}{N} + \frac{\sin(2L\sqrt{E})}{L} + \frac{2\sin(2\sqrt{E(L^2+N^2)})}{\sqrt{L^2+N^2}} - \frac{\pi}{L}\left[1 + J_0(2N\sqrt{E})\right]\right) \quad (3)$$

As in the case of 1D, this expression gives the correct result for an infinite system (only the first term in the rhs survives in this limit) and the agreement with the numerical results is again very satisfactory. The oscillations are present but the structure is far more complicated. Let us consider the case N=2L, which roughly corresponds to the actual dimensions of the Ag crystallites considered below[4]. The third term in the rhs of Eq. (3) approximately gives the same peak structure found in 1D, whereas the second promotes a splitting of each peak into two. The picture which is roughly valid for N larger than L is the following: a density of states showing a gross structure similar to that found in 1D with each peak splitted into approximately N/L peaks. This fine structure could not always be observed experimentally, depending upon resolution. As in 1D, peak separation depends on energy. It is interesting to analyse the density of states in a film ($L = \infty$). In this case only the first and second terms survive. The first term accounts for the density of states of an infinite system in 3D, whereas the second gives oscillations of the type discussed above. This result suggests that angularly integrated photoemission spectra of metallic films would show oscillations similar to those already found in angle-resolved spectra[1]. Finally we note that a linear dispersion relation gives a peak structure with peaks separated by a constant energy and a weight proportional to $E^2$.

Before entering into a more detailed analysis of the experimental results for Ag deposits on GaAs(110) several considerations are in order[2,3]: i) The observed oscillations sweap a rather large range of enegies, from the Fermi level down to around 4 eV; this is well below the top of the valence band of GaAs (less than 1 eV). ii) The coupling of the crystallites with the substrate should be rather small due to the significant lattice mismatch between the two solids[4] suggesting that these quantum well states weakly resonate with the valence states of GaAs. iii) The dispersion relation of Ag around the Fermi energy is linear in k. iv) In the triangular cross section of the crytallites the Ag atoms are distributed on a rectangular lattice[4] (($1\bar{1}0$) plane of the fcc lattice).

The above considerations, and the results for the free electron density of states in two and three dimensions, led us to carry out a calculation of the density states in triangular clusters of the square lattice by means of a tight binding model with one state per lattice site. This model can easily incorporate the rectangular lattice by taking different hopping integrals in the two directions of the square lattice; the three dimensional character of the crystallites can also be taken into account. Both, have been analysed and found not to change qualitatively the results herewith discussed, and their effects will be fully analysed elsewhere. Note that considering the fcc lattice and the sp character of the conduction electrons in silver, would strongly limit the size of the clusters that could be reachable. The model, on the other hand, gives a dispersion relation which, around the midband, is linear in k; note that in a metal (half filling) the midband coincides with the Fermi level. This overcomes the shortcoming of the free electron model discussed above which gives a quadratic dispersion relation for all energies. The triangular clusters considered here are bounded by the (01), (11) and ($\bar{1}$1) lines, the former being the one assumed to be in contact with the substrate. This model cannot of course account for all the details of the geometry of the Ag crystallites described in[4], but we think it takes into account their main features. Hereafter the unit of energy will be the hopping integral $t$ (note that typical values of $t$ adequate for normal metals would be in the range 1.0-2.0 eV).

The densities of states for triangular clusters having 51 and 77 atoms in their respective bases are shown in Fig. 2 (the total number of atoms in a given triangular cluster with $N_b$ atoms in its basis is $((N_b+1)/2)^2$). We first note that due to the linear character of the dispersion relation around the midband, the peaks are more regular in this energy region. Instead in the bottom and top of the band, where the dispersion relation is approximately quadratic, they are more irregularly distributed. Peak separation, on the other hand, is nearly 1.5 times larger in the smaller cluster. A fitting of the numerical results for clusters containing 576-9801 atoms gives a peak separation of $E_i - E_{i-1} = 1/(-0.24 + .19N_b)$ (with a correlation coefficient of 0.999). It should be noted that although these features of the density of states qualitatively agree with the experimental data a detailed comparison would require a full calculation including $s$ and $d$ atomic orbitals and clusters similar (in size and shape) to those grown in the experiment[2,4].

This model has also been used to calculate the angle resolved photoemission intensity. The latter has been approximated by the following expression[7],

$$f(E, k_{||}) = \sum_{i=1,N} |<e^{i\mathbf{kr}-d_s/2\lambda}|\mathbf{p}\cdot\mathbf{A}|\Psi_i>|^2$$
$$\delta(\omega - E_0 - W_0 + E_i) \qquad (4)$$

where $\omega$ is the photon energy, $W_0$ the work function, and E and $k_{||}$ the energy and momentum parallel to the surface of the photoemitted electron. In writing this expression, the outgoing electron has been described by a plane wave $exp(i\mathbf{kr})$ (with energy $E_0 = k^2 = (2\pi/a)^2(l^2+m^2)$, where a is the interatomic distance in the square lattice). The electron states within the crystallites $\Psi_i$ (energy $E_i$) are linear combinations of atomic orbitals which have been approximated by $\delta$-functions. $\lambda$ is the mean free path of the photoemitted electrons[7] and $d_s$ is the distance of a given atom to the surface of the crystallite, or the distance to either the (11) or the ($\bar{1}$1) lines (note that the cluster is assumed to be in touch with the substrate through the (01) line).

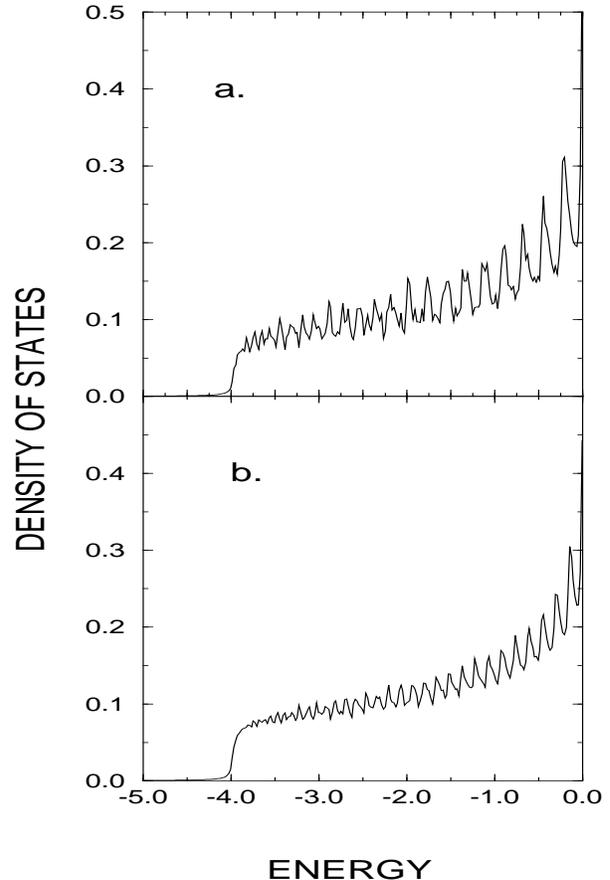

FIG. 2. Total density of states for two triangular clusters of the square lattice (defined by the (11), ($\bar{1}$1) and (01) lines) with 51 (a) and 77 (b) atoms in their bases, calculated within the tight binding approximation (one state per lattice site). Note that the density of states is symmetric with respect $E = 0$. The unit of energy is the hopping integral (see text).

The results for the photoemission intensity perpendicular to the (11) or ($\bar{1}$1) lines ($l = m$) are shown in Fig. 3. Figs. 3a and 3b shows the photoemitted intensity for $\omega - W_0 = 18$ and 12, respectively. The Fermi level is taken at the midband, $\lambda = 1.5a$ and $(2\pi/a)^2 = 10$. It is clearly noted that although the relative weights of the peaks change, their positions remain unchanged. This is a consequence of the reduced spatial dimensions of the cluster and is in agreement with the experimental data[2,3]. The effect of the emission direction is illustrated in Fig. 3c, where the results for $\omega - W_0 = 12$ and $m = 0.75l$ are plotted. We note important changes in the spectra, such as the dissapearance of some of the peaks, although the energies of the remaining peaks are not greatly modified. Again, these results qualitatively agree with the features observed in the experimental spectra.

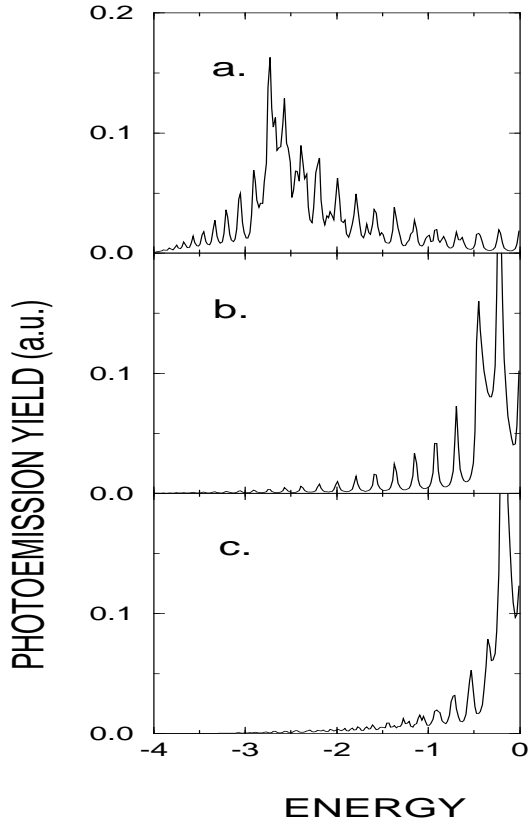

FIG. 3. Angle resolved photoemission intensity for three sets of parameters: a) $\omega - W_0 = 18$ (in units of the hopping integral) and emission perpendicular to the (11) or ($\bar{1}1$) lines ($l = m$), b) same as a) for $\omega - W_0 = 12$, and c) same as b) for $m = 0.75l$. Energies are referred to $E_F$. The unit of energy is the hopping integral (see text).

In summary, we have shown that the peak structure observed in photoelectron spectra of metallic deposits can only be unambiguously associated to single electron states if the deposit has a two dimensional character (such as in angle-resolved spectra of films). In one and zero dimensions the density of states shows peaks related to many electronic states, showing a fine structure (related to single electron states) which could eventually be resolved, depending upon experimental resolution, temperature and deposit dimensions. We have illustrated how the characteristics of the peak structure depend on the band dispersion in the energy region where they appear. A tight binding calculation of the electron states in triangular clusters of the square lattice has been used to discuss some of the features of the photoelectron spectra obtained for Ag deposits on GaAs(110).

## ACKNOWLEDGMENTS

We are greatful to M. Altarelli for helpful discussions. This work was supported in part by the spanish CICYT (grant MAT94-0058). V. Gasparian wishes to thank the "Ministerio de Educación y Ciencia" (Spain) for financial support (sabbatical grant SAB95-0085).

4